\documentclass[twocolumn,preprintnumbers,amsmath,amssymb]{revtex4}

\usepackage{graphicx}% Include figure files
\usepackage{dcolumn}% Align table columns on decimal point
\usepackage{bm}% bold math
\usepackage{color}
\usepackage{setspace}
\begin{document}

%\preprint{AIP/*****}

\title{Spin Seebeck effect in a polar antiferromagnet $\alpha$-Cu$_{2}$V$_{2}$O$_{7}$}% Force line breaks with \\

\author{Y. Shiomi$^{\, 1,2}$}
%\email{shiomi@imr.tohoku.ac.jp}
\author{R. Takashima$^{\, 3}$}
\author{D. Okuyama$^{\, 4}$}
\author{G. Gitgeatpong$^{\, 5,6,7}$}
\author{P. Piyawongwatthana$^{\, 5}$}
\author{K. Matan$^{\, 5,6}$}
\author{T. J. Sato$^{\, 4}$}
\author{E. Saitoh$^{\, 1,2,8,9}$}

%\affiliation{$^{1}$
%Department of Physics, Graduate School of Science, Tohoku University, Sendai, 980-8578, Japan}
\affiliation{$^{1}$
Institute for Materials Research, Tohoku University, Sendai 980-8577, Japan }
\affiliation{$^{2}$
Spin Quantum Rectification Project, ERATO, Japan Science and Technology Agency, Aoba-ku, Sendai 980-8577, Japan}
\affiliation{$^{3}$
Department of Physics, Kyoto University, Kyoto 606-8502, Japan}
\affiliation{$^{4}$
IMRAM, Tohoku University, Sendai 980-8577, Japan 
}
\affiliation{$^{5}$
Department of Physics, Faculty of Science, Mahidol University, Bangkok 10400, Thailand
}
\affiliation{$^{6}$
ThEP, Commission of Higher Education, Bangkok 10400, Thailand
}
\affiliation{$^{7}$
Department of Physics, Faculty of Science and Technology, Phranakhon Rajabhat University, Bangkok 10220, Thailand
}
\affiliation{$^{8}$
Advanced Institute for Materials Research, Tohoku University, Sendai 980-8577, Japan
}
\affiliation{$^{9}$
Advanced Science Research Center, Japan Atomic Energy Agency, Tokai 319-1195, Japan
}

\date{\today}% It is always \today, today,

\begin{abstract}
We have studied the longitudinal spin Seebeck effect in a polar antiferromagnet $\alpha$-Cu$_{2}$V$_{2}$O$_{7}$ in contact with a Pt film. Below the antiferromagnetic transition temperature of $\alpha$-Cu$_{2}$V$_{2}$O$_{7}$, spin Seebeck voltages whose magnetic field dependence is similar to that reported in antiferromagnetic MnF$_{2}$$\mid$Pt bilayers are observed. Though a small weak-ferromagnetic moment appears owing to the Dzyaloshinskii-Moriya interaction in $\alpha$-Cu$_{2}$V$_{2}$O$_{7}$, the magnetic field dependence of spin Seebeck voltages is found to be irrelevant to the weak ferromagnetic moments. The dependences of the spin Seebeck voltages on magnetic fields and temperature are analyzed by a magnon spin current theory. The numerical calculation of spin Seebeck voltages using magnetic parameters of $\alpha$-Cu$_{2}$V$_{2}$O$_{7}$ determined by previous neutron scattering studies reveals that the magnetic-field and temperature dependences of the spin Seebeck voltages for $\alpha$-Cu$_{2}$V$_{2}$O$_{7}$$\mid$Pt are governed by the changes in magnon lifetimes with magnetic fields and temperature.
\end{abstract}

%\pacs{72.25.-b, 75.76.+j, 74.25.Fy, 74.25.Qt, 74.72.Bk, 72.20.Pa}% PACS, the Physics and Astronomy
\maketitle

%\section*{Introduction}

Low-dimensional quantum magnets have attracted much attention in condensed matter physics for many decades \cite{bonner, sachdev}. It is known that in a one-dimensional antiferromagnetic system, long-range ordering is absent even at zero temperature \cite{bethe}, leading to exotic magnetic ground states, {\it e.g.} quantum spin liquid states \cite{diep}. Spin excitations in spin liquid states are spinons, which are regarded as spin-$1/2$ excitations as opposed to spin-$1$ excitations of magnons. Very recently, spin currents carried by spinons were demonstrated experimentally using the spin Seebeck effect (SSE) in Sr$_{2}$CuO$_{3}$$\mid$Pt systems \cite{hirobe}. Unusual magnetic properties of low-dimensional quantum magnets are intriguing for the search of new spin current effects.   
\par

\begin{figure}[t]
\begin{center}
\includegraphics[width=8cm]{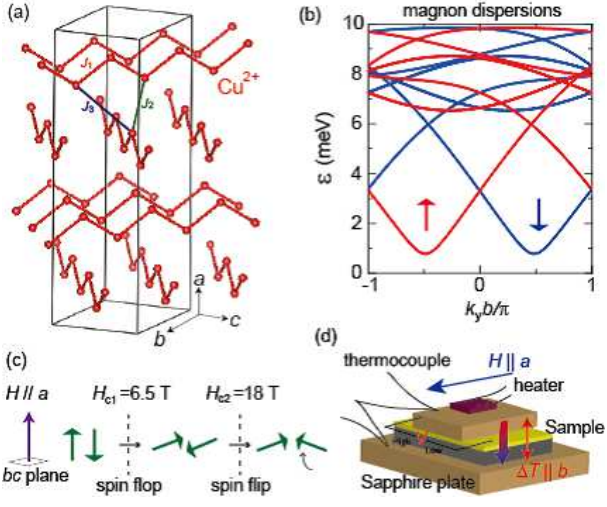}
\caption{(a) Cu ions form a pair of almost perpendicular zigzag chains. Spin exchange interactions ($J_{1}$, $J_{2}$, and $J_{3}$) are also shown. (b) Magnon dispersions of $16$ branches calculated using magnetic parameters determined by the neutron scattering study [\onlinecite{GG-prl2017}]. (c) Two magnetic transitions observed in $H||a$. The first transition occurring at $6.5$ T at $1.4$ K is a spin-flop transition, and the second transition at $18$ T at $1.4$ K is a spin-flip transition of canted moments.  (d) Experimental setup of the longitudinal SSE. The sample, $\alpha$-Cu$_{2}$V$_{2}$O$_{7}$$\mid$Pt, is sandwiched by two sapphire plates. The temperature difference ($\Delta T$) is applied using a chip-resistor heater, and measured using a couple of thermocouples.   } 
\label{fig1}
\end{center}
\end{figure}

Among various compounds, a copper divanadates $\alpha$-Cu$_{2}$V$_{2}$O$_{7}$ is a low-dimensional antiferromagnetic spin-$1/2$ system with fascinating magnetic properties. The oxide compound Cu$_{2}$V$_{2}$O$_{7}$ crystallizes in dichromate structure with three different polymorphs, {\it i.e.} $\alpha$, $\beta$, and $\gamma$ phases \cite{bhowal}. The structures of the $\beta$ and $\gamma$ phases are centrosymmetric, while the $\alpha$ phase possesses a noncentrosymmetric crystal structure with a polar point group ($mm2$) \cite{bhowal, Sanningrahi-prb2015}. The space group of $\alpha$-Cu$_{2}$V$_{2}$O$_{7}$ is $Fdd2$ with lattice constants of $a= 20.645$ \AA, $b=8.383$ \AA, and $c=6.442$ \AA \cite{GG-prb2015}. As illustrated in Fig. \ref{fig1}(a), all Cu$^{2+}$ ions form two sets of almost perpendicular zigzag chains \cite{bhowal, Sanningrahi-prb2015, GG-prb2015, GG-prb2017}.
\par

Magnetic properties of $\alpha$-Cu$_{2}$V$_{2}$O$_{7}$ are governed by Cu$^{2+}$ $S=1/2$ spins, since V$^{5+}$ ions are nonmagnetic. In the proposed spin Hamiltonian \cite{GG-prl2017}, there are three important terms to explain the magnetic properties of $\alpha$-Cu$_{2}$V$_{2}$O$_{7}$. The first term is isotropic exchange interactions. In the zigzag spin chains, Cu$^{2+}$ spins interact with their nearest neighbors. Since the magnetic interactions between nearest ($J_{1}$), second-nearest ($J_{2}$), and third-nearest ($J_{3}$) neighbors are all antiferromagnetic \cite{bhowal, Sanningrahi-prb2015, GG-prb2015, GG-prb2017, GG-prl2017}, $\alpha$-Cu$_{2}$V$_{2}$O$_{7}$ exhibits antiferromagnetism. The second term is an anisotropic exchange interaction, which arises from the multiorbital correlation effect \cite{GG-prl2017}. The third term is the Dzyaloshinskii-Moriya (DM) interaction, which results from the absence of the inversion center between the nearest-neighbor spins \cite{GG-prl2017}. Because of the DM interaction, ``Rashba-type" splitting of lowest magnon bands is realized \cite{GG-prl2017, hayami, cheng}, as shown in Fig. \ref{fig1}(b). Nonreciprocal magnons were clearly observed very recently in inelastic neutron scattering measurements \cite{GG-prl2017}.
\par

In this manuscript, we study a thermal generation effect of spin currents, {\it i.e.} the longitudinal SSE \cite{uchida}, in antiferromagnetic $\alpha$-Cu$_{2}$V$_{2}$O$_{7}$. The SSE in antiferromagnetic insulators is a hot topic in the recent spin-caloritronics field \cite{seki, wu}. The generated spin current in $\alpha$-Cu$_{2}$V$_{2}$O$_{7}$ is detected by an electric voltage in an attached Pt film, which results from the conversion from the spin current into charge current by means of the inverse spin Hall effect. The clear observation of the SSE in antiferromagnetic $\alpha$-Cu$_{2}$V$_{2}$O$_{7}$ allows us to test the theoretical models proposed for the antiferromagnetic SSE \cite{ohnuma, rezende}. Using the magnon spin current theory \cite{rezende} combined with magnetic parameters determined by previous neutron experiments for $\alpha$-Cu$_{2}$V$_{2}$O$_{7}$ \cite{GG-prb2015, GG-prb2017, GG-prl2017}, we demonstrated that temperature and magnetic-field dependences of magnon lifetimes are important for the antiferromagnetic SSE in $\alpha$-Cu$_{2}$V$_{2}$O$_{7}$$\mid$Pt systems.   
\par

%\section*{Experiments}

Single crystals of $\alpha$-Cu$_{2}$V$_{2}$O$_{7}$ were grown by a vertical Bridgman method following the previous reports \cite{GG-prb2015, GG-prb2017, GG-prl2017}. A crushed crystal was cut into a cuboid with the size of $4 \times 2 \times 0.6$ mm$^{3}$. The widest plane is a crystallographic $ac$ plane; the longest side ($4$ mm long) is parallel to the $c$ axis. On the $ac$ plane, a $5$-nm-thick Pt film was sputtered in an Ar atmosphere at ambient temperature. For the measurement of the SSE, the $\alpha$-Cu$_{2}$V$_{2}$O$_{7}$$\mid$Pt sample was sandwiched by two sapphire plates \cite{shiomi-1, shiomi-2}, as illustrated in Fig. \ref{fig1}(d); on one plate, a $1$-k${\rm \Omega}$ chip resistor was fixed with GE varnish to apply temperature gradient along the $b$ axis of $\alpha$-Cu$_{2}$V$_{2}$O$_{7}$. The temperature difference, $\Delta T$, arising between two sapphire plates was measured using a couple of type-E thermocouples. The thermoelectric voltage was measured along the $c$ axis under external magnetic fields ($H$) applied along the $a$ axis. The measurement was performed in a cryogen-free superconducting magnet at each $H$ between $-9$ T and $9$ T; $H$ was first decreased from $9$ T to $-9$ T, then increased backed to $9$ T. The antisymmetric contribution of the thermoelectric voltage spectra, $\{ V_{\rm raw}(H)-V_{\rm raw}(-H)\}/2$, is defined as the spin Seebeck voltage $V$. To compare the spin Seebeck voltages with magnetization data of $\alpha$-Cu$_{2}$V$_{2}$O$_{7}$, the magnetization for the same $\alpha$-Cu$_{2}$V$_{2}$O$_{7}$ sample was measured under $H||a$ with a vibrating sample magnetometer in a superconducting magnet.     
\par

\begin{figure}[t]
\begin{center}
\includegraphics[width=8cm]{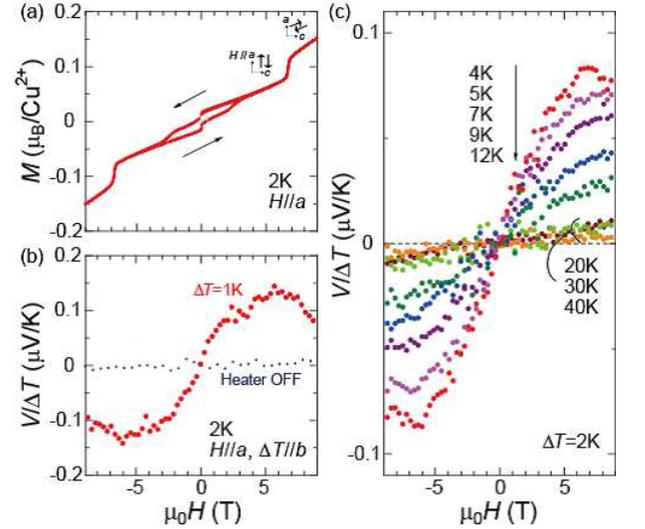}
\caption{(a) Magnetization, $M$, as a function of magnetic field ($H$) with $H||a$ at $2$ K. (b) $H$ dependence of the spin Seebeck voltage divided by the temperature difference, $V/\Delta T$, at $2$ K. Here, $H$ is applied along the $a$ axis, and $\Delta T$ along the $b$ axis. The background voltage measured in the heater-off condition is also shown. (c) $V/\Delta T$ as a function of $H$ with $H||a$ at different temperatures from $4$ K to $40$ K. The magnitude of the applied $\Delta T$ is fixed at $2$ K. } 
\label{fig2}
\end{center}
\end{figure}

Figure \ref{fig2}(a) shows the $H$ dependence of the magnetization, $M$, for $\alpha$-Cu$_{2}$V$_{2}$O$_{7}$ at $2$ K. Here, the magnetization was measured in the $H$ range from $-9$ T to $9$ T. It has been reported \cite{GG-prb2015, GG-prb2017, GG-prl2017} that in the low $H$ regime below $6.5$ T, the Cu$^{2+}$ spins align antiparallel with their nearest and next-nearest neighbors, and the majority of the spin component points along the $a$ axis. As shown in Fig. \ref{fig2}(a), $M$ increases almost in proportion to $H$, which is consistent with the antiferromagnetic alignment of Cu$^{2+}$ spins. However, in the low-$H$ regime below $3$ T, a weak ferromagnetic moment with a hysteresis loop was observed in the magnetization curve. The weak ferromagnetic moment whose magnitude is almost $0.01$ ${\rm \mu_{B}}$/Cu$^{2+}$ at $H \approx 0$ is ascribed to small field-induced canting due to the DM interaction \cite{bhowal, Sanningrahi-prb2015, GG-prb2015, GG-prb2017, GG-prl2017, BC}. When $H$ increases up to $6.5$ T, $M$ abruptly increases from $0.07$ to $0.11$ ${\rm \mu_{B}}$/Cu$^{2+}$. This transition is a spin-flop transition of the Cu$^{2+}$ spins \cite{GG-prb2017} [Fig. \ref{fig1}(c)]. The competition between the exchange energy and Zeeman energy forces the spins to minimize the total energy by flopping altogether into the $bc$ plane making the direction of the majority of the spin component perpendicular to $H$ \cite{GG-prb2017}. Above $6.5$ T, owing to the presence of the $a$ component of the DM vector, the spin components in the $bc$ plane form a helical structure with the helical axis along the $a$ axis \cite{GG-prb2017}. 
\par

In Fig. \ref{fig2}(b), the spin Seebeck voltage divided by the temperature difference, $V/\Delta T$, measured at $2$ K is presented. When the heater is turned on so that the temperature difference $\Delta T$ is set to be $1$ K, the clear spin Seebeck voltage is observed. With increasing $H$, the spin Seebeck voltage increases monotonically up to $6$ T. Importantly, the $H$ dependence of $V/\Delta T$ is totally different from that of $M$ in Fig. \ref{fig2}(a), whereas it was reported that the magnitude of the antiferromagnetic spin Seebeck voltage is always proportional to magnetization in Cr$_{2}$O$_{3}$$\mid$Pt \cite{seki}. The weak ferromagnetic moment and the hysteresis loop were not observed in the spin Seebeck voltage in $\alpha$-Cu$_{2}$V$_{2}$O$_{7}$$\mid$Pt. By contrast, the nonlinear $H$ dependence of $V/\Delta T$ for $\alpha$-Cu$_{2}$V$_{2}$O$_{7}$$\mid$Pt is found to be similar to that reported for the antiferromagnetic SSE in MnF$_{2}$$\mid$Pt below the spin-flop transition field \cite{wu}. Above $6$ T, $V/\Delta T$ tends to decrease, which seemingly contradicts with the sharp increase in the spin Seebeck voltage above the spin flop transitions in Cr$_{2}$O$_{3}$$\mid$Pt \cite{seki} and MnF$_{2}$$\mid$Pt \cite{wu}. A difference in the magnetic state in the spin-flopped phase for $\alpha$-Cu$_{2}$V$_{2}$O$_{7}$ from MnF$_{2}$ and Cr$_{2}$O$_{3}$ is that the helical spin structure is realized owing to the DM interaction in $\alpha$-Cu$_{2}$V$_{2}$O$_{7}$ in contrast to typical spin-flop transitions in MnF$_{2}$ and Cr$_{2}$O$_{3}$. The helical spin structure stabilized by the DM interaction might have a low efficiency for spin current generation, as reported for a skyrmion insulator Cu$_{2}$OSeO$_{3}$ \cite{hirobe-skyrmion}. It is noted that we confirmed that $V/\Delta T$ is almost zero in the entire $H$ regime when the heater is turned off ($\Delta T \approx 0$), as shown in Fig. \ref{fig2}(b).    
\par

The $H$ dependence of $V/\Delta T$ at various temperatures is shown in Fig. \ref{fig2}(c). Below $12$ K, the clear spin Seebeck voltage which increases with increasing $H$ strength is observed. The suppression of the spin Seebeck voltage in the high-$H$ spin-flopped phase is observed also at $4$ K. As temperature ($T$) increases from $4$ K, $|V/\Delta T|$ decreases monotonically, and is hardly discerned above $20$ K. Above $20$ K, small $H$-linear voltages were only observed, which can be ascribed to the normal Nernst effect in Pt films \cite{kikkawa}. The sign of the spin Seebeck voltage in $\alpha$-Cu$_{2}$V$_{2}$O$_{7}$$\mid$Pt is the same as that of the normal Nernst effect in Pt, which is consistent with the SSE reported for various magnets \cite{uchida-review}. 
\par

\begin{figure}[t]
\begin{center}
\includegraphics[width=8cm]{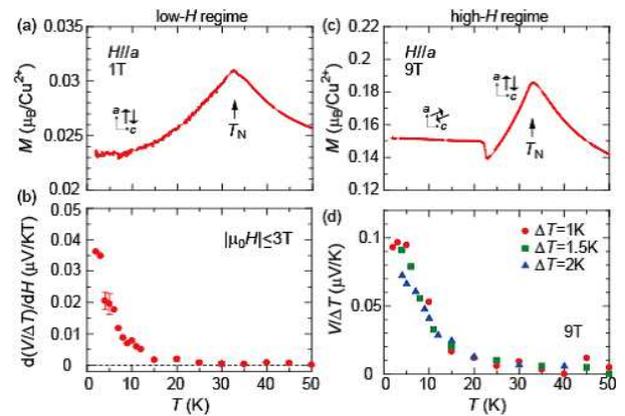}
\caption{(a) Temperature ($T$) dependences of the magnetization ($M$) [(a) and (c)] and the spin Seebeck signal [(b) and (d)] in the low-field [(a) and (b)] and high-field regimes [(c) and (d)]. In (b), the slopes of the linear fits to $V/\Delta T$ in the $H$ range between $-3$ T and $3$ T are shown. In (d), $T$ dependences of $V/\Delta T$ measured at different $\Delta T$ values ($1$ K, $1.5$ K, and $2$ K) are shown. } 
\label{fig3}
\end{center}
\end{figure}

In Fig. \ref{fig3}, $T$ dependences of $M$ and $V/\Delta T$ are compared in the low-$H$ regime [(a) and (b)] and high-$H$ regime [(c) and (d)]. Figure \ref{fig3}(a) shows $T$ dependence of $M$ measured in a field-cooling scan at $1$ T. The magnetization exhibits a cusp at $\approx$$33$ K, which corresponds to the antiferromagnetic transition of $\alpha$-Cu$_{2}$V$_{2}$O$_{7}$. The reported N\'eel temperature $T_{N}$ is $33.4$ K \cite{GG-prb2015}, very close to the present transition temperature. The decrease in $M$ below $T_{N}$ is consistent with the antiparallel alignment of the majority of Cu$^{2+}$ spins along the $a$ axis. 
\par

In Fig. \ref{fig3}(b), the $T$ dependence of the spin Seebeck signal in the low-$H$ regime is presented. Here, the $H$ dependence of $V/\Delta T$ in the range of $-3$ T $\leq$$\mu_{0}H$$\leq$ $+3$ T is fitted with a linear function of $H$, and the slope ${\rm d} (V/\Delta T)/{\rm d} (\mu_{0}H) $ is plotted. The spin Seebeck signal is almost zero above $20$ K, but sharply increases with decreasing $T$ below $20$ K. The strong suppression of the spin Seebeck signal at $20$ K well below $T_{N}$ suggests that the $T$ dependence of the SSE depends not only on magnetic parameters but on transport parameters. The monotonic increase in the spin Seebeck voltage with decreasing $T$ down to $2$ K is similar to that reported in MnF$_{2}$$\mid$Pt bilayers; the spin Seebeck voltage at $1$ T for MnF$_{2}$$\mid$Pt monotonically increases with decreasing $T$ down to $3$ K \cite{wu}.      
\par

In the high-$H$ regime above $6.5$ T, the magnetization shows a complex $T$ dependence below $T_{N}$, as shown in Fig. \ref{fig3}(c). Below $T_{N}$, $M$ first decreases with decreasing $T$ consistent with the antiparallel spin alignment along the $a$ axis, but becomes almost independent of $T$ below $\approx$$23$ K. In the low-$T$ range below $\approx$$23$ K, the Cu$^{2+}$ spins flop into the $bc$ plane, and the helical spin structure is realized. In the spin-flopped phase, the spin Seebeck voltage tends to be suppressed, as shown at $2$ K [Fig. \ref{fig2}(b)] and $4$ K [Fig. \ref{fig2}(c)]. Nevertheless, $V/\Delta T$ measured at $9$ T increases monotonically with decreasing $T$ below $T_{N}$, as shown in Fig. \ref{fig3}(d). The $T$ dependence of $V/\Delta T$ is less sensitive to the spin flop transition than that of $M$.
\par

\begin{figure}[t]
\begin{center}
\includegraphics[width=8cm]{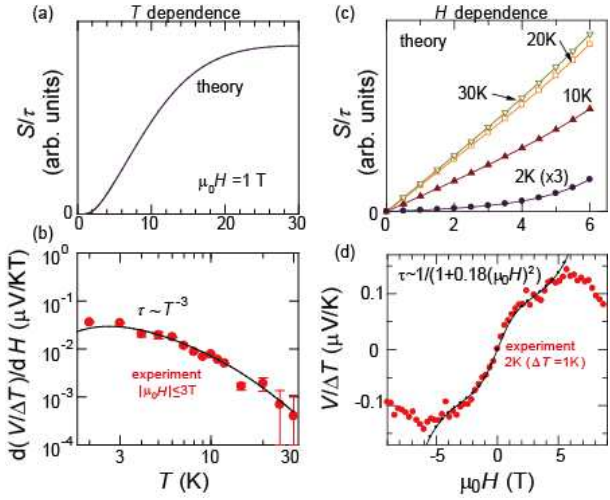}
\caption{(a) Temperature ($T$) dependence of the calculated spin Seebeck coefficient ($S$) divided by the magnon lifetime ($\tau$) at $1$ T. (b) A log-log plot of the low-field spin Seebeck signal already shown in Fig. \ref{fig3}(b). The experimental data is fitted using a theoretical curve obtained from the calculation shown in (a) and assuming $\tau \propto T^{-3}$. (c) Magnetic field ($H$) dependence of the calculated $S/\tau$ at $2$ K, $10$ K, $20$ K, and $30$ K. (d) $H$ dependence of the spin Seebeck voltage divided by the temperature difference, $V/\Delta T$, at $2$ K, which was already shown in Fig. \ref{fig2}(b). The experimental data in the low-$H$ range below the spin flop transition field ($\approx 6.5$ T) is fitted using the calculated curve in (c) with the assumption that $\tau \propto 1/(1+ 0.18(\mu_{0}H)^2)$. } 
\label{fig4}
\end{center}
\end{figure}

The $T$ and $H$ dependences of the spin Seebeck voltage in $\alpha$-Cu$_{2}$V$_{2}$O$_{7}$$\mid$Pt are similar to those reported in MnF$_{2}$$\mid$Pt below the spin-flop transition field \cite{wu}. The $H$ dependence of $V/\Delta T$ which is irrelevant to the weak ferromagnetic moments [Figs. \ref{fig2}(b) and \ref{fig2}(c)] indicates that antiferromagnetic magnons are responsible for the SSE. The monotonic increase in $V/\Delta T$ with decreasing $T$ down to the lowest $T$ [Figs. \ref{fig3}(b) and \ref{fig3}(c)] is unlikely to be correlated with the $T$ dependence of thermal conductivity, since the thermal conductivity for bulk single crystals usually shows a maximum at $10$-$30$ K \cite{ziman}. Though longitudinal spin Seebeck systems {\it e.g.} Y$_{3}$Fe$_{5}$O$_{12}$$\mid$Pt have a correlation between the size of the spin Seebeck signal and the thermal conductivity \cite{Guo, kikkawa-2015, jin, boona, iguchi}, the phonon thermal transport seems not important for the longitudinal SSE in $\alpha$-Cu$_{2}$V$_{2}$O$_{7}$$\mid$Pt. Recently, Rezende {\it et al} explained the antiferromagnetic SSE by magnon spin currents driven by temperature gradient \cite{rezende}. Following the magnon spin current theory \cite{rezende}, we will analyze the SSE observed in $\alpha$-Cu$_{2}$V$_{2}$O$_{7}$$\mid$Pt. 
\par

From the Boltzmann equation with the relaxation time approximation, the magnon spin current due to the temperature gradient ${\vec J}_{\nabla T}$ is given by \cite{rezende}
\begin{equation}
\vec{J}_{\nabla T}=
-\hbar
\int \frac{d^3 k}{(2\pi)^3}
\sum_{\sigma, n}
\sigma 
 \vec v_{\sigma n k} 
(\vec v_{\sigma n k} \cdot \vec \nabla T)\ \tau
\frac{\partial n^{0}_{\sigma n k}}{\partial T},
\end{equation}
to the lowest order of the temperature gradient. Here, $\vec v_{\sigma n k} $ is the group velocity of magnons with spin $\sigma $ ($\sigma =\pm 1$)  and momentum $\vec k$ for the band $n$ [Fig. \ref{fig1}(b)], and it is defined  by $\vec v_{\sigma n k} = \partial \varepsilon_{\sigma n k } /(\hbar \partial \vec{k})$ with $\varepsilon_{\sigma n k }$ being the energy for the magnon mode. $n^{0}_{\sigma n k}= 1/(e^{\varepsilon_{\sigma n k }/k_B T}-1) $ is the Bose-Einstein distribution function. $\tau=\tau(T, H)$ is the relaxation time (lifetime) of magnons that depends on the temperature and the magnetic field, where we have neglected its spin, momentum and energy band dependence. The response function (spin Seebeck coefficient) $S$ that is proportional to the experimental $V/\Delta T$ thereby reads 
\begin{equation}
\frac{V}{\Delta T} \propto S = -\hbar \tau \int \frac{d^3 k}{(2\pi)^3} 
\sum_{\sigma, n}
\sigma
\frac{\partial n_{\sigma n k}^{0}}{\partial T}  v_{\sigma n k}^{y\ \ 2}.
\end{equation}
%It is noted that nonreciprocal transport due to $(\Delta T)^{2}$ is neglected, since all the momentum modes are thermally excited in the SSE, and since magnon group velocities of up-spin and down-spin modes are unchanged with respect to the band split by the DM interaction.
\par 

Temperature dependence of $S$ originates from the magnon relaxation time $\tau$ and the distribution function $\partial n_{\sigma n k}^{0}/\partial T$; the latter part is important especially at lower temperatures than the $T$ scale of the magnon band gap. In Fig. \ref{fig4}(a), $T$ dependence of $S/\tau$ for $\mu_{0}H =1$ T is calculated from eq.(2) using the magnon dispersions in Fig. \ref{fig1}(b) \cite{GG-prl2017}. The $T$ dependence of $S/\tau$ is relatively small above $20$ K, and exponentially decreases to zero below $\approx$$15$ K. Since the experimental $V/\Delta T$ increases with decreasing $T$ [Fig. \ref{fig3}] in contrast to the monotonic decrease in $S/\tau$ [Fig. \ref{fig4}(a)], $\tau$ is responsible for the experimental $T$ dependence of the SSE. In Fig. \ref{fig4}(b), the low-$H$ data of the SSE shown in Fig. \ref{fig3}(b) is fitted using the calculated response function $S$ in Fig. \ref{fig4}(a). Here, $\tau$ is a control parameter for the fit. As shown in Fig. \ref{fig4}(b), the response function $S$ assuming $\tau \propto T^{-3}$ well explains the experimental $T$ dependence below $30$ K. The $T$ dependence of $\tau$ is attributable to 4-magnon relaxation processes \cite{harris}, and it was reported that $\tau \propto T^{-3}$ for antiferromagnetic MnF$_{2}$ \cite{rezende, Bayrakci}, the same as the present case. The calculated peak $T$ is $2.5$ K, consistent with the monotonic increase in $V/\Delta T$ down to $3$ K [Fig. \ref{fig3}(b)]. 
\par

In the presence of external magnetic fields, the up-spin (down-spin) magnon modes shift downwards (upwards) owing to the Zeeman effect. The energy shift due to the Zeeman effect gives rise to the $H$ dependence of the SSE. In Fig. \ref{fig4}(c), the $H$ dependence of $S/\tau$ is calculated at $2$ K, $10$ K, $20$ K, and $30$ K. The $H$ dependence of $S/\tau$ is almost linear above $10$ K, and super-linear at $2$ K. In the experimental $H$ dependence of $V/\Delta T$ in Fig. \ref{fig2}, the $H$ dependence is almost linear at the high-$T$ range, but clearly convex upward at $2$ K [Fig. \ref{fig2}(b)]. The nonlinear $H$ dependence of $V/\Delta T$ observed at $2$ K is not explained by the calculated $H$ dependence of $S/\tau$ in Fig. \ref{fig4}(c), and thus ascribed to the $H$ dependence of $\tau$. In the case of MnF$_{2}$$\mid$Pt, Rezende {\it et al.} pointed that the $H$ dependence of $\tau$ is more significant at lower temperatures, and that the scattering rate ($= 1/\tau$) increases with $H$ in proportion to power series of $H$ \cite{rezende}. For $\alpha$-Cu$_{2}$V$_{2}$O$_{7}$$\mid$Pt, if we assume that $1/\tau \propto 1+ 0.18(\mu_{0}H)^2$, the $H$ dependence of the response function $S$ calculated in Fig. \ref{fig4}(c) well explains the experimental $H$ dependence of $V/\Delta T$ below the spin flop transtion, as shown in Fig. \ref{fig4}(d).   
\par

In summary, the longitudinal SSE was studied in $\alpha$-Cu$_{2}$V$_{2}$O$_{7}$$\mid$Pt. The observed inverse spin Hall voltage induced by the SSE is not proportional to the magnetization, but similar to the antiferromagnetic SSE reported recently for MnF$_{2}$$\mid$Pt systems. Using the magnon spin-current theory combined with magnetic parameters determined by previous neutron scattering studies, we discussed the temperature and magnetic-field dependences of the spin Seebeck voltage, and clarified that the change in the magnon scattering rate with temperature and magnetic fields plays an important role for the antiferromagnetic SSE in $\alpha$-Cu$_{2}$V$_{2}$O$_{7}$$\mid$Pt systems.    
\par

%acknowledgement
The authors thank T. Kikkawa for critical comments on the manuscript. This work was supported by JST ERATO ``Spin Quantum Rectification Project" (JPMJER1402), JSPS KAKENHI (No. 17H04806, No. JP16H00977, No. 16K13827, No. 17K18744, and No. 15J01700), and MEXT (Innovative Area ``Nano Spin Conversion Science" (No. 26103005)). Work at IMRAM is partly supported by the research program ``Dynamic alliance for open innovation bridging human, environment, and materials".
\par

%
%\clearpage

%\clearpage

%\clearpage

%\clearpage

\end{document}